\documentclass{article}
\usepackage{amsmath,epsfig,graphicx,booktabs,multirow,cleveref, subcaption,graphicx,stfloats}
\usepackage[preprint]{spconfa4}

\copyrightnotice{ 978-1-6654-3864-3/21/\$31.00~\copyright 2021 IEEE}

\let\OLDthebibliography\thebibliography
\renewcommand\thebibliography[1]{
  \OLDthebibliography{#1}
  \setlength{\parskip}{0pt}
  \setlength{\itemsep}{0pt plus 0.3ex}
}

\begin{document}\sloppy

\def\x{{\mathbf x}}
\def\L{{\cal L}}

\title{Visual Analysis Motivated Rate-Distortion Model for Image Coding}
%
\name{Zhimeng Huang$^{\ast}$, Chuanmin Jia$^{\ast}$, Shanshe Wang$^{\ast}$$^{\dagger}$, Siwei Ma$^{\ast}$$^{\dagger}$}
\address{$^{\ast}$Institute of Digital Media, Peking University, Beijing, China\\
$^{\dagger}$Information Technology R\&D Innovation Center of Peking University, Shaoxing, China\\
Email: $\{$zmhuang, cmjia, sswang, swma$\}$@pku.edu.cn
\thanks{This work was supported in part by the National Natural Science Foundation of China (62072008, 61931014), the China Postdoctoral Science Foundation (2020M680238), PKU-Baidu Fund (2019BD003) and High-performance Computing Platform of Peking University, which are gratefully acknowledged.
}
}
\maketitle

\begin{abstract}
	Optimized for pixel fidelity metrics, images compressed by existing image codec are facing systematic challenges when used for visual analysis tasks, especially under low-bitrate coding. This paper proposes a visual analysis-motivated rate-distortion model for Versatile Video Coding (VVC) intra compression. The proposed model has two major contributions, a novel rate allocation strategy and a new distortion measurement model. We first propose the region of interest for machine (ROIM) to evaluate the degree of importance for each coding tree unit (CTU) in visual analysis. Then, a novel CTU-level bit allocation model is proposed based on ROIM and the local texture characteristics of each CTU. After an in-depth analysis of multiple distortion models, a visual analysis friendly distortion criteria is subsequently proposed by extracting deep feature of each coding unit (CU). To alleviate the problem of lacking spatial context information when calculating the distortion of each CU, we finally propose a multi-scale feature distortion (MSFD) metric using different neighboring pixels by weighting the extracted deep features in each scale. Extensive experimental results show that the proposed scheme could achieve up to 28.17\% bitrate saving under the same analysis performance among several typical visual analysis tasks such as image classification, object detection, and semantic segmentation.  
\end{abstract}
\begin{keywords}
	VVC intra, bit allocation, rate distortion optimization, convolutional neural network 
\end{keywords}
\section{Introduction}
\begin{figure}
	\centering
	\subfloat[Top1-Acurracy - Rate curve]
	{
		\begin{minipage}[t]{0.24\textwidth}
			\centering
			\includegraphics[width=0.98\textwidth]{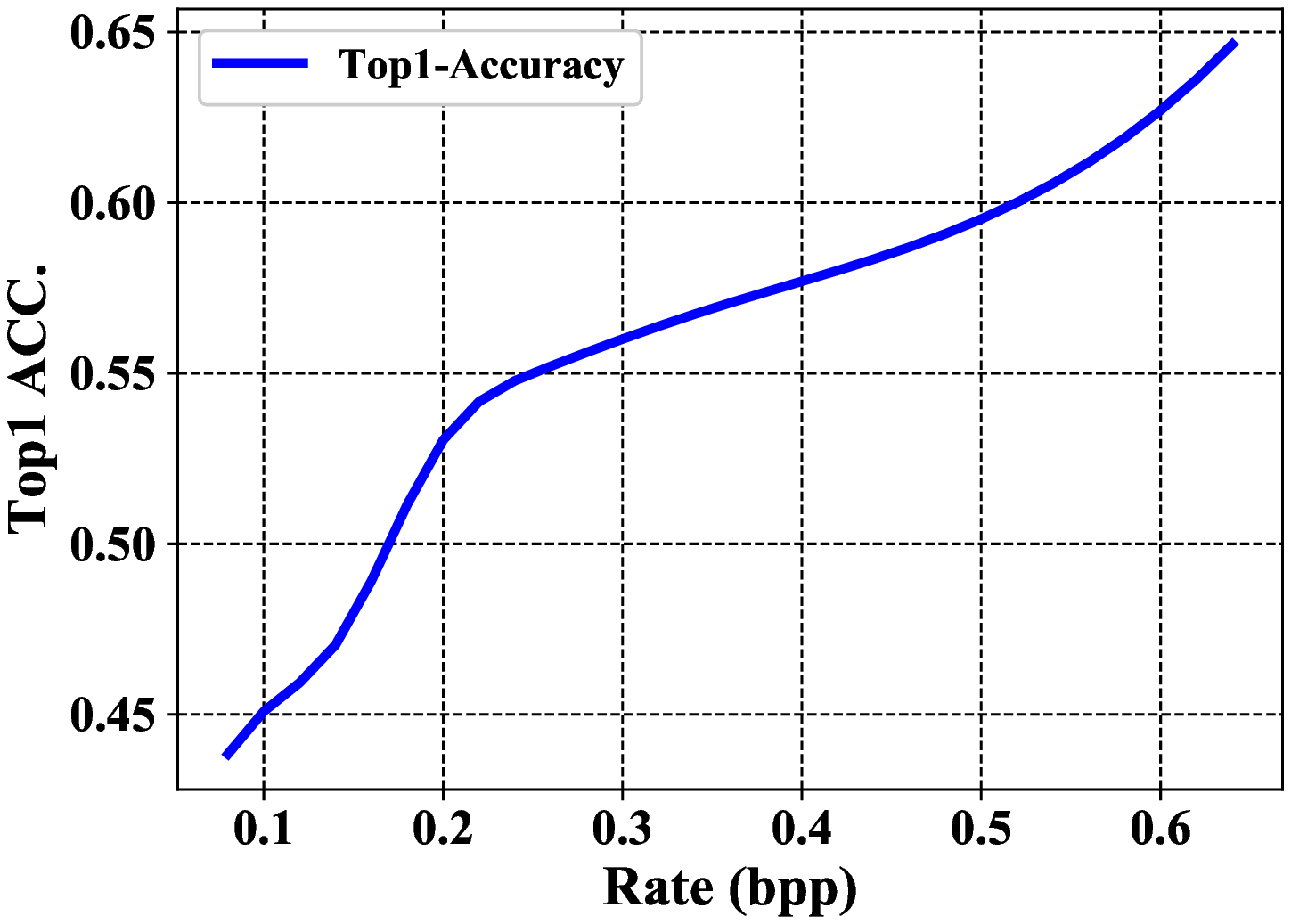}
		\end{minipage}
		\label{fig:top1}
	}
	\subfloat[Top5-Acurracy - Rate curve]
	{
		\begin{minipage}[t]{0.24\textwidth}
			\centering
			\includegraphics[width=0.98\textwidth]{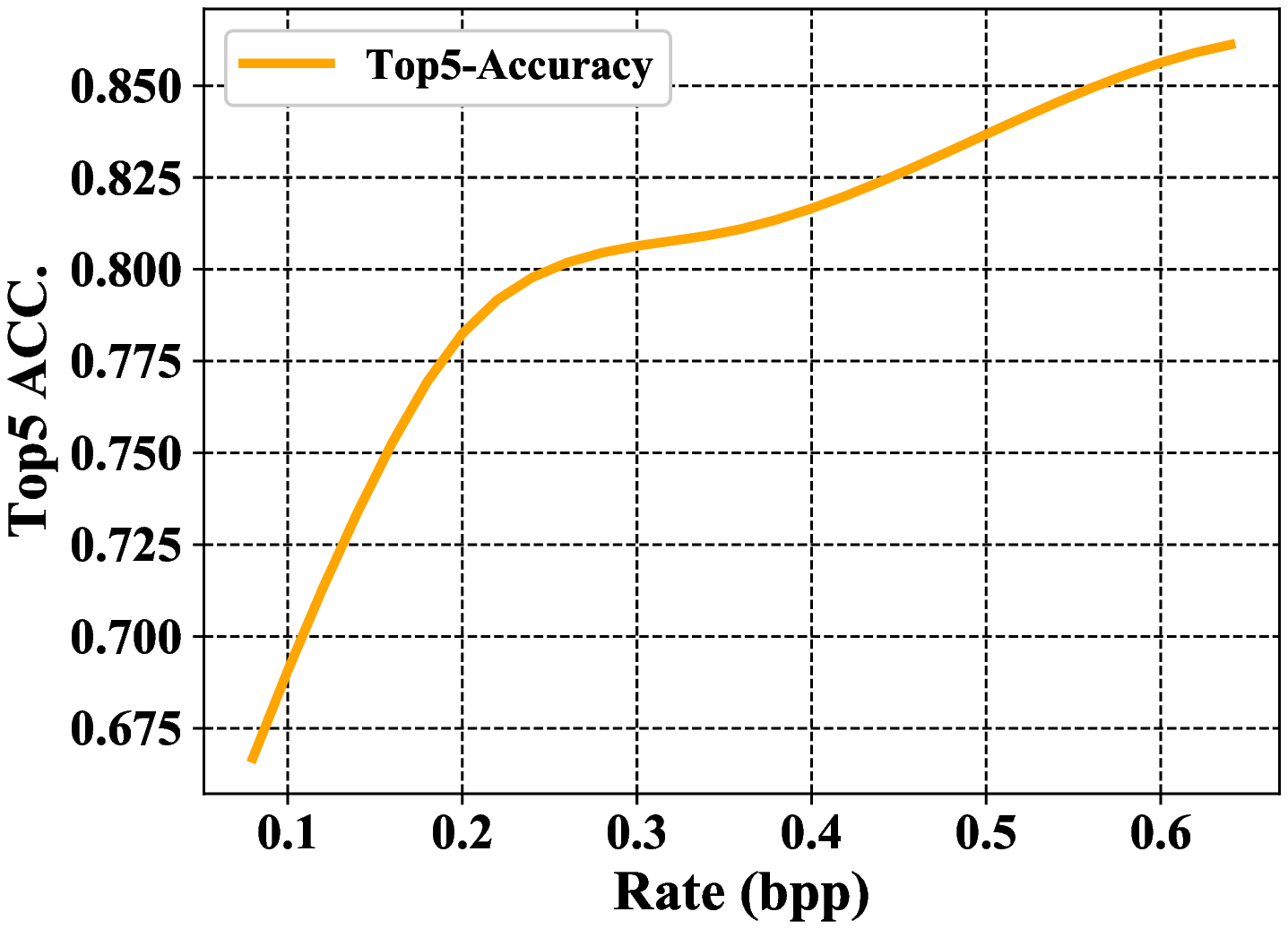}
		\end{minipage}
		\label{fig:top2}
	}
	\caption{Performance of VGG-19 network dealing with ImageNet dataset in low bitrate on image classification task.}
	\label{fig:decrease}
\end{figure}

With the rapid development of visual analysis and image/video understanding, plenty of intelligent applications (e.g., image classification, person re-identification, medical image diagnosis) arise. Lots of deep learning based machine intelligence algorithms have shown fantastic performance on these visual analysis tasks. For example, on image classification task, the top-5 accuracy of ResNet-50\cite{he2016deep} model on ImageNet\cite{deng2009imagenet} is over 97\%. However, the input images of these networks are usually uncompressed or compressed with high quality. Experimental results reveal that the analysis performance on these tasks severely degrades when dealing with images encoded in low bitrate. As shown in Fig.~\ref{fig:decrease}, the top-5 accuracy of image classification is only 68\% when the bits per pixel (bpp) equaling to 0.1. This phenomenon indicates that existing coding methods are not effective enough when dealing with visual analysis tasks, especially in low bitrate.
\par Rate-distortion optimization (RDO) plays an important role in improving performance in image coding. According to \cite{chen2019learning}, types of distortions are classified into three categories: pixel level distortion, perceptual distortion for human vision, and semantic distortion for machine vision. The most commonly used pixel fidelity metric is mean square error (MSE). It has been adopted as the optimization objective in both traditional codecs\cite{zhang2019recent, sullivan2012overview,CFP} and deep learning-based end-to-end compression approaches\cite{toderici2017full}\cite{balle2018variational}. However, the pixel level distortion metric has its limitations. As mentioned in \cite{wang2009mean}, many images with intolerant noises have a negligible MSE. To deal with the mismatching between MSE and human viewing experiences, a large number of distortion metrics in perceptual fidelity are proposed\cite{egiazarian2006new,li2011visual}. These new metrics promote lots of new RDO approaches to improve perceptual quality\cite{hadizadeh2013saliency,khanna2015perceptual,oh2012video,zhu2018spatiotemporal}. However, both distortion metrics in pixel level and perceptual level perform poorly in visual analysis tasks. As image/video coding for machine vision has become an emerging topic in recent years, how to utilize characteristics of the machine vision system (MVS) to encode  visual analysis-friendly image/video becomes an essential challenge. Chen et al.\cite{chen2019learning} propose a Regionally Adaptive Pooling (RAP) module which could apply Generative Adversarial Network (GAN) as metric directly in image compression scheme to improve performance on facial analysis tasks. Shi et al.\cite{shi2020reinforced} combine reinforcement learning with High Efficiency Video Coding (HEVC) to determine QP based on prior knowledge of the target visual task. However, it is still a huge challenge to define a general semantic distortion for visual analysis tasks. 
\begin{figure*}[t]
	\centering
	\includegraphics[width=0.98\textwidth]{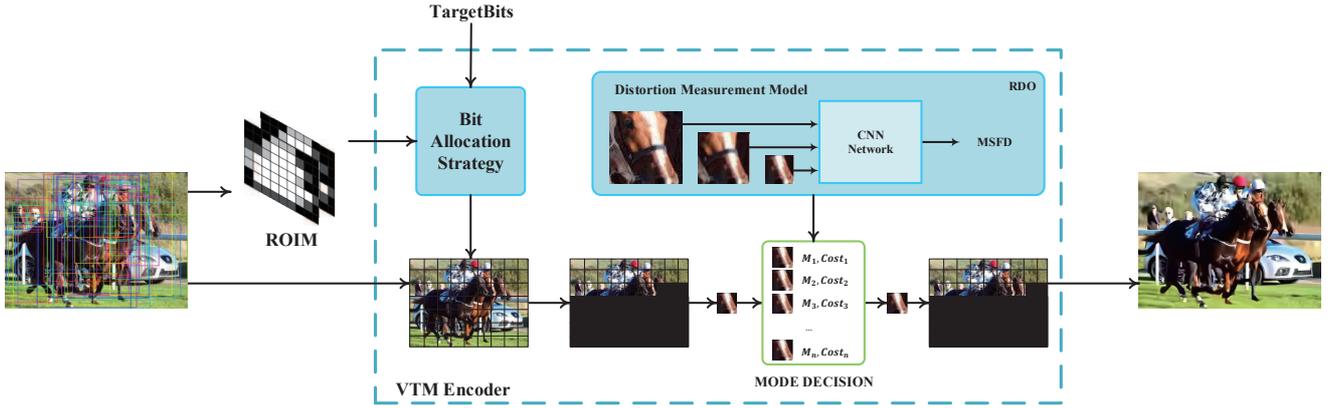}
	\caption{Illustration of the proposed framework. ROIM indicates the region of interest for machine. $M_i$ denotes different coding modes in mode decision and $cost_i$ means the rd-cost calculated by RDO module for each coding mode. MSFD is the multi-scale feature distortion proposed in our optimization model.}
	\label{Framework}
\end{figure*} 
\par In this paper, we propose a novel visual analysis-motivated RDO model for VVC intra compression. The framework of the proposed model is shown in Fig.~\ref{Framework}. Our contributions can be summarized as follows:
\begin{itemize}
	\item For bit allocation, a new concept named ROIM is proposed to evaluate the degree of the importance for each coding tree unit (CTU) on visual analysis tasks. We calculate a normalized density of bounding boxes to estimate the ROIM for each CTU. The CTU-level bit allocation strategy is guided by the ROIM and local texture information. 
	\item By an in-depth analysis among multiple CNN based feature extraction methods to measure the distortion in visual analysis, VGG-11 without last pooling and fully connected layer is adopted and utilized as our distortion measurement in the proposed approach.
	\item To deal with lacking spatial context information, a multiscale feature extracting method is proposed to increase the context information for the coding unit (CU). The proposed model utilizes the reconstructed pixels as reference to expand the receptive field.
\end{itemize}

\section{Proposed method}
In this section, we introduce the details of the proposed method, including the bit allocation strategy and the distortion measurement model.
\subsection{Bit Allocation Strategy}
\subsubsection{ROIM Generation}
\par
\begin{figure}[h]
	\centering
	\subfloat[case 1]
	{
		\begin{minipage}[t]{0.24\textwidth}
			\centering
			\includegraphics[width=0.98\textwidth]{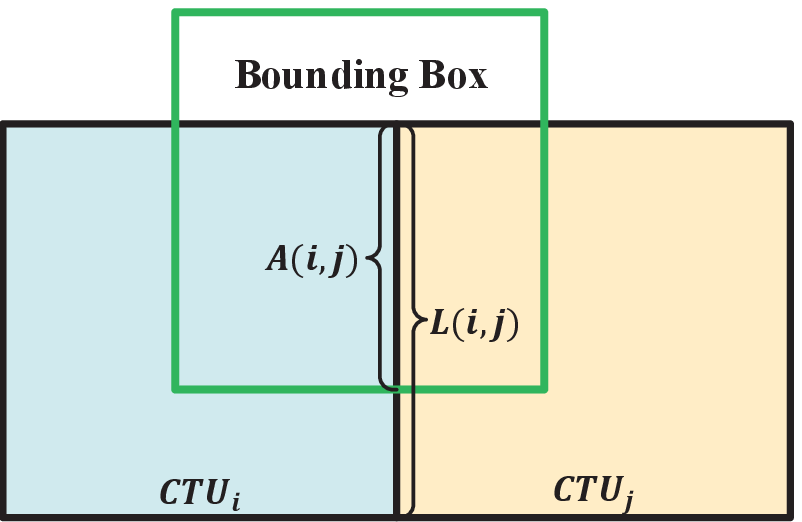}
		\end{minipage}
	}
	\subfloat[case 2]
	{
		\begin{minipage}[t]{0.24\textwidth}
			\centering
			\includegraphics[width=0.49\textwidth]{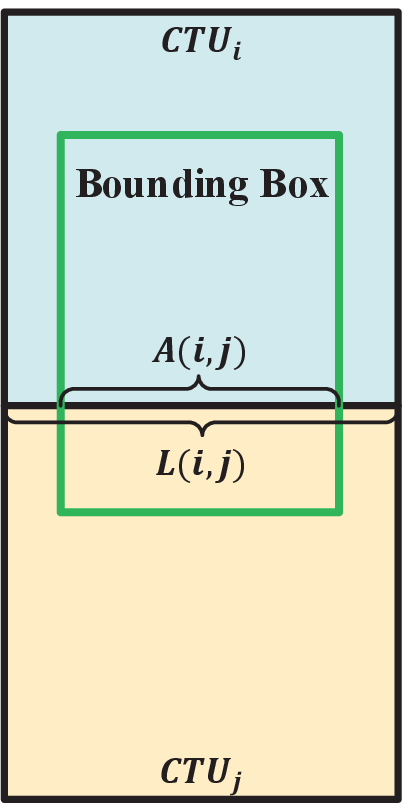}
		\end{minipage}
	}
	\caption{Two cases while calculating the $M_c(i,j)$. In case(a), $CTU_i$ and $CTU_j$ are horizontally adjacent and in case(b), they are vertically adjacent.}
	\label{fig:Connectivity}
\end{figure}

 In VTM, the basic unit of bit allocation is CTU. Following the reference software, the proposed model also chooses CTU as the unit of ROIM. The ROIM module consists of two parts: $M_i$ and $M_c$. $M_i$ reveals the degree of importance for each CTU and $M_c$ indicates the degree of connectivity for adjacent CTUs. The network for ROIM is built on the top of a pre-trained RPN. The RPN could generate a set of possible bounding boxes named $B$ before the NMS (Non-Maximum Suppression) procedure. After the extraction, the degree of importance for $CTU_k$ is calculated by:
\begin{equation}\label{eq:imp}
	M_i\left(k\right) = \cfrac{\sum\limits_{b\in{B}}f\left(k\bigcap b\right)}{\max\limits_{k\in{S_C}}\left(\sum\limits_{b\in{B}}f\left(k\bigcap b\right)\right)},k\in{S_C},
\end{equation}
where $S_C$ denotes the index set of CTUs. Function $f(A)$ is used for calculating the number of pixels in region $A$. The degree of connectivity is evaluated to limit the QP relationships of adjacent CTUs, which can be modeled by:
\begin{equation}\label{eq:cnt}
	M_{c}\left(i,j\right) = 
	\cfrac{A(i,j)}{L(i,j)},
\end{equation}	
where $L(i,j)$ represents the length of the boundary adjacent to $CTU_i$ and $CTU_j$, and $A(i,j)$ denotes the length of the boundary adjacent to $CTU_i$ and $CTU_j$ covered by bounding boxes Fig.~\ref{fig:Connectivity} gives an example of the calculation of function $A(i,j)$ and $L(i,j)$.

\subsubsection{Bit Allocation}
In VTM, the CTU-level bit allocation aims at allocating the bits based on the texture information. $Target_{CTU_i}$ denotes the target bit allocated for $CTUi$, which is calculated by:
\begin{equation}
	Target_{CTU_i} = \frac{(Target_{Pic}-Bits_{coded})*SATD_{i}}{\sum\limits_{j\in{S_U}}SATD_{j}},
\end{equation}
where $Target_{Pic}$ represents the target bit of the whole frame. $i$ and $j$ are indexes of CTUs. $Bits_{coded}$ denotes the bits that have been used for the current frame. $S_U$ indicates the indexes of CTUs that have not been coded. The $SATD_{i}$ is the sum of absolute transformed difference for $CTUi$. $SATD_{i}$ could estimate the complexity of the content roughly. As a result, the quality of every CTU in the coded image could be more consistent. Motivated by visual analysis tasks and local context information, we propose a novel bit allocation method:
\begin{equation}\label{eq:cost}
	Cost_k = \left(SATD_{k}/3+\alpha*M_i\left(k\right)\right),
\end{equation}
\begin{equation}\label{eq:SHAD}
	Target_{CTU_k} = \frac{(Target_{Pic}-Bits_{coded})*Cost_k}{\sum\limits_{j\in{S_U}}Cost_j},
\end{equation}
where $\alpha$ is the trade-off between $SATD$ and $M_i$ since the orders of magnitudes between them are totally different.
\subsubsection{QP constraint}
\newcounter{TempEqCnt}
\setcounter{TempEqCnt}{\value{equation}}
\setcounter{equation}{9}
\begin{figure*}[hb]
	\begin{equation}
		QP_{CUEst_i}=\left\{
		\begin{aligned}
			Clip(QP_{CTU_k}-2,QP_{CTU_k}+2,QP_{CUEst_i}),\quad if M_{c}\left(i,k\right)> 0.7\\
			Clip(QP_{CTU_k}-9,QP_{CTU_k}+9,QP_{CUEst_i}),\quad if M_{c}\left(i,k\right)\leq 0.7\\
		\end{aligned}
		\right.
		\label{QP5}
	\end{equation}
\end{figure*}
\setcounter{equation}{\value{TempEqCnt}}
In principle, QP estimation is a part of RDO. However, considering that multiple-QP optimization will significantly increase encoding complexity, QP estimation is separated from the RDO module in VTM. Therefore a QP limitation for every CTU is proposed to keep the reconstructed image consistent\cite{li2014lambda}. The QP estimation procedure should satisfy Eqn.~(\ref{QP1}) and Eqn.~(\ref{QP2}). 

\begin{equation}
	QP_{CUEst} = Clip(QP_{CU}-1,QP_{CU}+1,QP_{CUEst})\label{QP1},
\end{equation}
\begin{equation}
	QP_{CUEst} = Clip(QP_{Pic}-2,QP_{Pic}+2,QP_{CUEst})\label{QP2},
\end{equation}
where $QP_{pic}$ represents the QP of the picture and $QP_{CU}$ is the average QP of the coded CTUs. The function $Clip$ means the third parameter is limited between the first and the second parameter to prevent blocking effects. However, blocking effects caused by QP estimation usually will not reduce the performance on the visual analysis tasks. Therefore, a new QP limitation strategy based on the degree of connectivity for adjacent CTUs is proposed. We first find a $CTUk$ as the limitation of the coding $CTUi$:
\begin{equation}
	k = \arg\max_j{M_c(i,j)},
	\label{calk}
\end{equation}
where $i$ denotes the index of the current CTU. Finally, the QP of the coding CTU is calculated by Eqn.~(\ref{QP5}), in which $QP_{CTU_k}$ denotes the QP of $CTUk$.
Moreover, the limitation between the $QP_{CUEst}$ and $QP_{Pic}$ is removed. The proposed limitation could keep the reconstructed region of bounding boxes consistent and remove the limitation between the object areas and the background areas. 

\subsection{Rate Distortion Optimization}
In this section, a novel RDO model based on CNN extracting features is proposed. The proposed algorithm mainly deals with two problems: the choice of the neural network and how to fully utilize the limited information in CU. Different from the pixel fidelity, the similarity of features in the proposed model is calculated by:
\begin{equation}
	FD = (1-\frac{RecF\cdot OriF}{\left\|RecF\right\|\left\|OriF\right\|}),
\end{equation}
where $RecF$ denotes the deep features extracted from the reconstructed picture and $OriF$ denotes the deep features extracted from the original picture.
\subsubsection{Distortion Measurement Analysis} 
We first analyze the performance of four common CNN based feature distortion on semantic segmentation task. To avoid the influence of the bit allocation strategy, only the rate-distortion calculation part of VVC test model (VTM)\cite{VTM10.1} is modified.  The experiments are conducted among ResNet18, ResNet34, VGG-11, and VGG-16 without the last pooling layer and fully connected layers. These models are all pre-trained by ImageNet, which is enormous and manifold enough. 100 images from the valid set of COCO-2014 dataset are randomly selected to compare the speed of the network and the performance on semantic segmentation task. With the bpp aligning, we compare the mean average precision with confidence equalling to 0.50 (mAP@50) on semantic segmentation task and time cost compared with VGG-11 model. As Table \ref{tab:comparision} shown, the mAP@50 of VGG network based optimization model is better than ResNet based model. As for the two VGG based models, the VGG-16 shows negligible performance improvement even with a nearly 10\% increase in time consumption. Since the traditional codec uses RDO module very frequently, the smaller model VGG-11 is selected as the feature extractor.   
\begin{table}[h]
	\begin{center}
		\caption{Comparison of mAP@50 on segmentation task and speed compared with VGG-11 among four common CNN model.} 
		\label{tab:comparision}
		\begin{tabular}{|c|c|c|c|}
			\hline
			Module & BPP & mAP@50&Time Cost\\
			\hline
			ResNet18&0.107&0.26&78\%\\
			ResNet34&0.109&0.28&102\%\\
			VGG-11&0.111&0.31&100\%\\
			VGG-16&0.114&0.32&110\%\\
			\hline
		\end{tabular}
	\end{center}
\end{table}
\subsubsection{Multi-Scale Feature Distortion}
The partition in traditional codec will generate lots of small blocks. Moreover, the receptive fields of these blocks are too small to extract discriminative deep features. To alleviate the problem of lacking the context information when calculating the distortion of each CU, the framework utilizes a multi-scale window to extract a series of context information from coded CU to make the features more discriminative. As the pixels on the top and left have been reconstructed, the proposed model utilizes them as reference pixels.
\begin{figure}[h]
	\centering
	\includegraphics[width=3.5in]{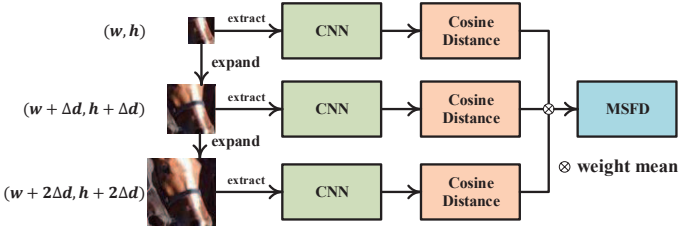}
	\caption{Multi-scale structure to extract MSFD}
	\label{fig:multi-scale}
\end{figure} 
\newline
Fig.~\ref{fig:multi-scale} shows the multi-scale structure extracting deep features with different sizes. $\Delta d$ denotes the extra dims to expand. By calculating the weighted cosine distance between the feature of reconstructed CU and that of original CU, a visual analysis motivated distortion is put forward:
\setcounter{equation}{10}
\begin{equation}
	MSFD = \sum_i{w_i*FD_i}*W*H,
	\label{MSFD}
\end{equation}
where $FD_i$ is the cosine distance calculated with the multi-scale window $i$, $w_i$ are the weights of different windows. $W$ and $H$ represent the width and height of the coding block.
\par Although the multi-scale window structure could increase the context information for CUs with small size. It is still difficult to extract deep features from CUs with extremely small sizes such as 4x4. So we use the max-value of cosine distance to estimate the $FD$ of the blocks with height or width smaller than 16. However, this approximation may cause severe pixel level distortion. To balance the quality of reconstructed regions, the traditional MSE loss is added to distortion as an amendment. The MSE loss is calculated by:
\begin{equation}\label{MSE}
	MSE = \frac{1}{N}\sum_i{(Rec_i-Org_i)^2}, 
\end{equation}
where $N$ is the number of the pixels of the CU, $Rec_i$ and $Org_i$ are the pixel values in the reconstructed unit and original unit. 
All in all, the final distortion is calculated by:
\begin{equation}\label{D}
	D =  MSFD + \beta*MSE.
\end{equation} 
Considering that the max-value of cosine distance is 2 which is much smaller than MSE, we utilize a trade-off parameter $\beta$ to adjust the orders of magnitudes of two kinds of distortions.
\section{Experiments and Results}
\subsection{Visual Analysis Tasks and Datasets}
To prove the universality of the framework, three different visual tasks are employed for the experiment including image classification, object detection and semantic segmentation. For image classification, we randomly select 1000 images in the ImageNet valid dataset and use VGG-19 image classification network to test the accuracy of the top-1 and the top-5. For object detection, we randomly select 1000 images in VOC-2007 valid dataset and use YOLOV3 to test the mAP@50. For semantic segmentation, we randomly select 1000 images in COCO-2014 valid dataset and use Mask r-cnn to test the mAP@50. The calculation method of mAP refers to the dataset VOC. The reason why subsets are adopted is that the complexity of the model results in a long experimental time. As for the ROIM, it is possible to generate ROIM in the bit allocation part and QP estimation part of the codec. However, the generation of ROIM for a certain image is independent of the configurations for codec. So the ROIM of each test image is pre-processed off-line.
\subsection{Configuration of Experiments}
The proposed approach is implemented on the top of VTM 10.1. 3000 images are employed for testing with QP equalling to 40,42,44,46. We take these images encoded by VTM 10.1 as the anchor. The reason why we only compare the performance of visual tasks with high QP is that the differences of performance in high bitrate are negligible. The configuration for the encoder is the common test condition of all intra with rate control enabled. The target bitrate of the proposed approach is adjusted to align the bitrates of the anchor. The practical bitrates rather than the target bitrates are counted for further comparison. 
\par As for the hyper-parameters in the model, $\alpha$ in the Eqn.~(\ref{eq:SHAD}) is 10000. The $\Delta d$ in Section 2.2 is 8. The size of multi-window is 3, and the weighted parameter $w_i$ is \{4, 2, 1\}, respectively.
\begin{figure*}[!htb]
	\centering
	\subfloat[]
	{
		\begin{minipage}[t]{0.22\textwidth}
			\centering
			\includegraphics[width=0.98\textwidth]{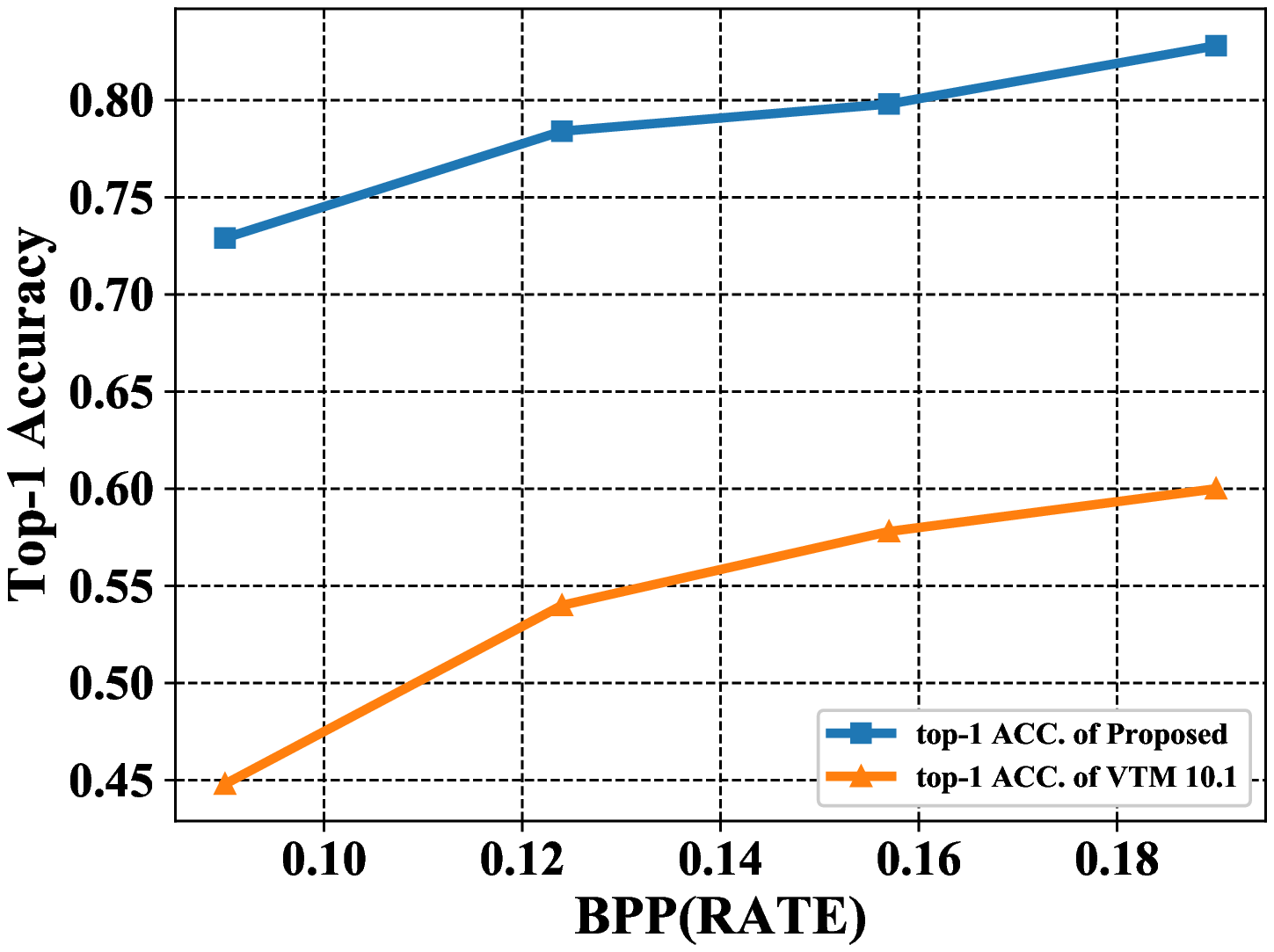}
		\end{minipage}
		\label{fig:classification_top1_curve}
	}
	\subfloat[]
	{
		\begin{minipage}[t]{0.22\textwidth}
			\centering
			\includegraphics[width=0.98\textwidth]{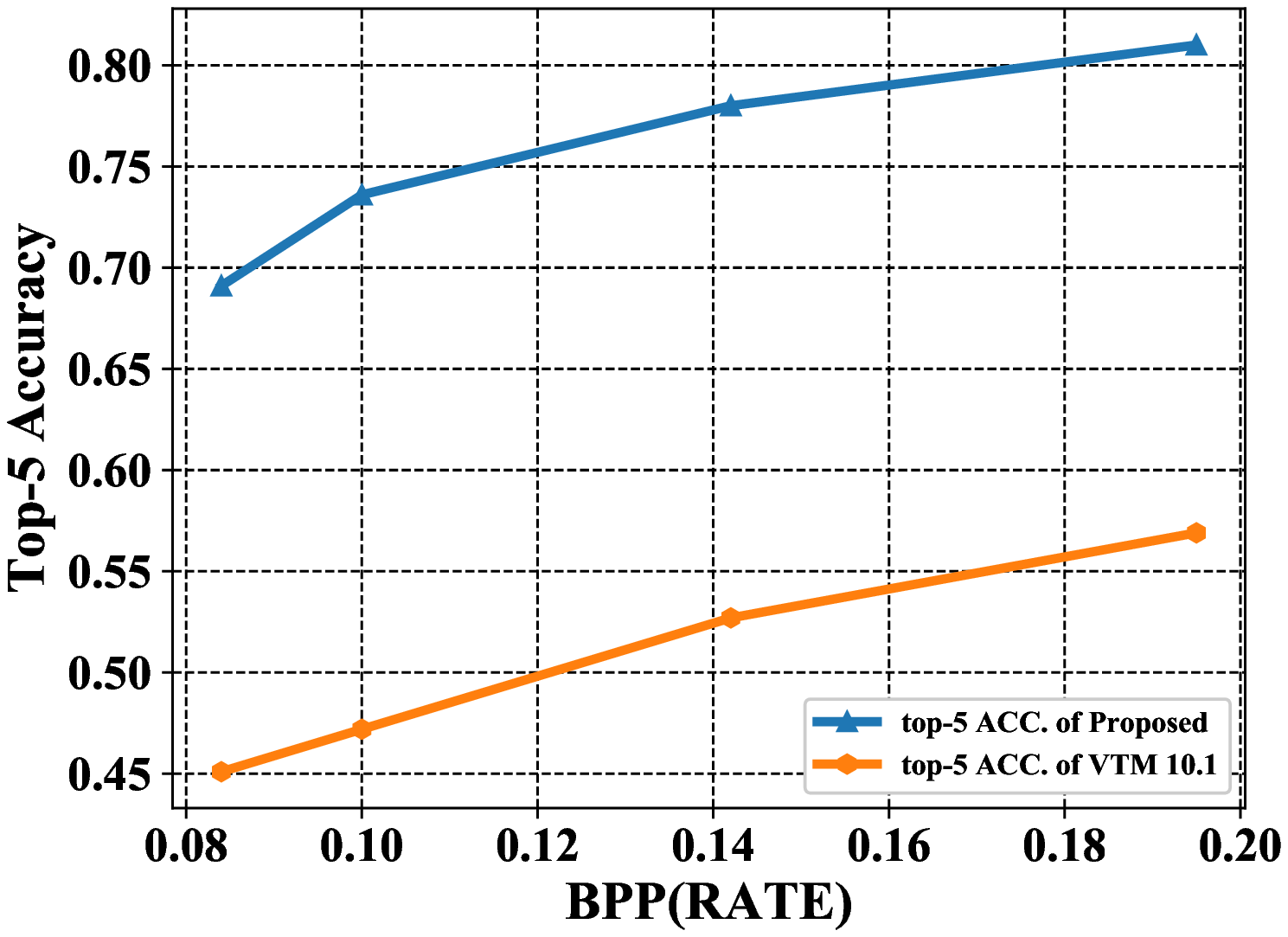}
		\end{minipage}
		\label{fig:classification_top5_curve}
	}
	\subfloat[]
	{
		\begin{minipage}[t]{0.22\textwidth}
			\centering
			\includegraphics[width=0.98\textwidth]{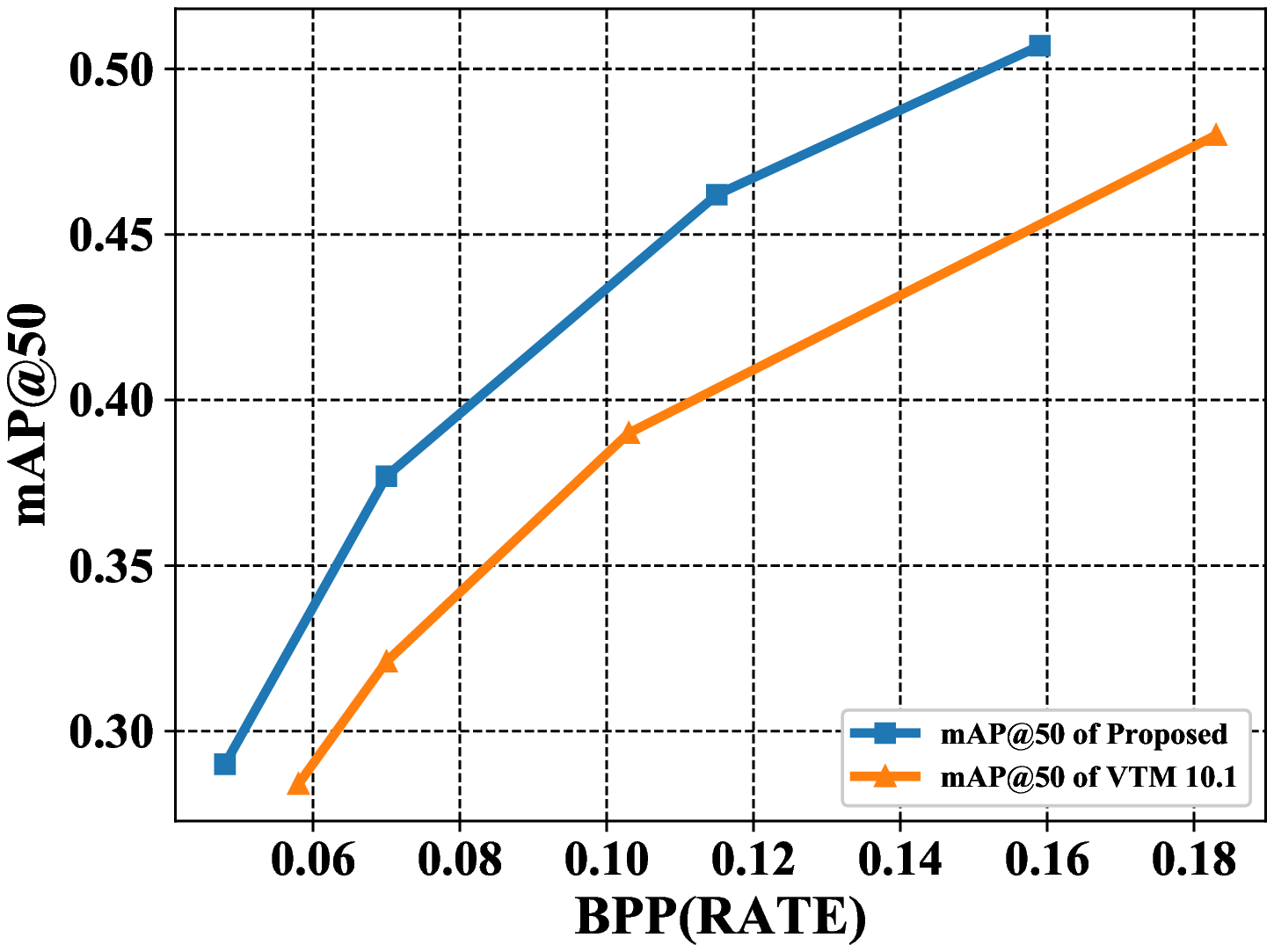}
		\end{minipage}
		\label{fig:detection_curve}
	}
	\subfloat[]
	{
		\begin{minipage}[t]{0.22\textwidth}
			\centering
			\includegraphics[width=0.98\textwidth]{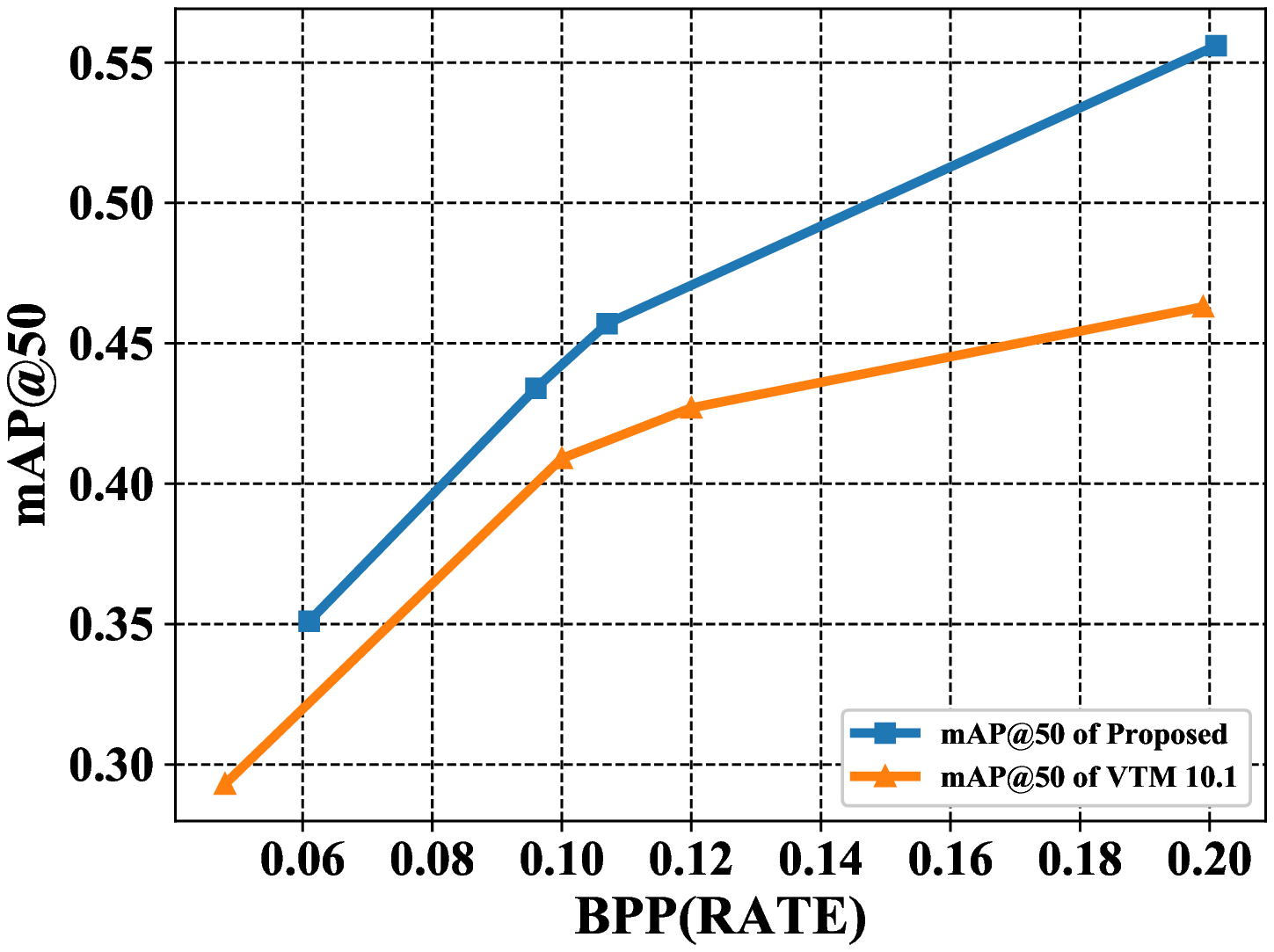}
		\end{minipage}
		\label{fig:segmentation_curve}
	}
	\caption{(a) shows the comparison of top-1 accuracy between the proposed model and VTM 10.1 on image classification task on ImageNet dataset; (b) shows the comparison of top-1 accuracy between the proposed model and VTM 10.1 on image classification task on ImageNet dataset; (c) shows the comparison of mAP@50 between the proposed model and VTM 10.1 on object detection task on VOC-2007 dataset; (d) shows the comparison of mAP@50 between the proposed model and VTM 10.1 on semantic segmentation task on COCO2014 dataset.}
	\label{fig:curves}
\end{figure*}
\begin{figure*}[h]
	\centering
	\includegraphics[width=0.95\textwidth]{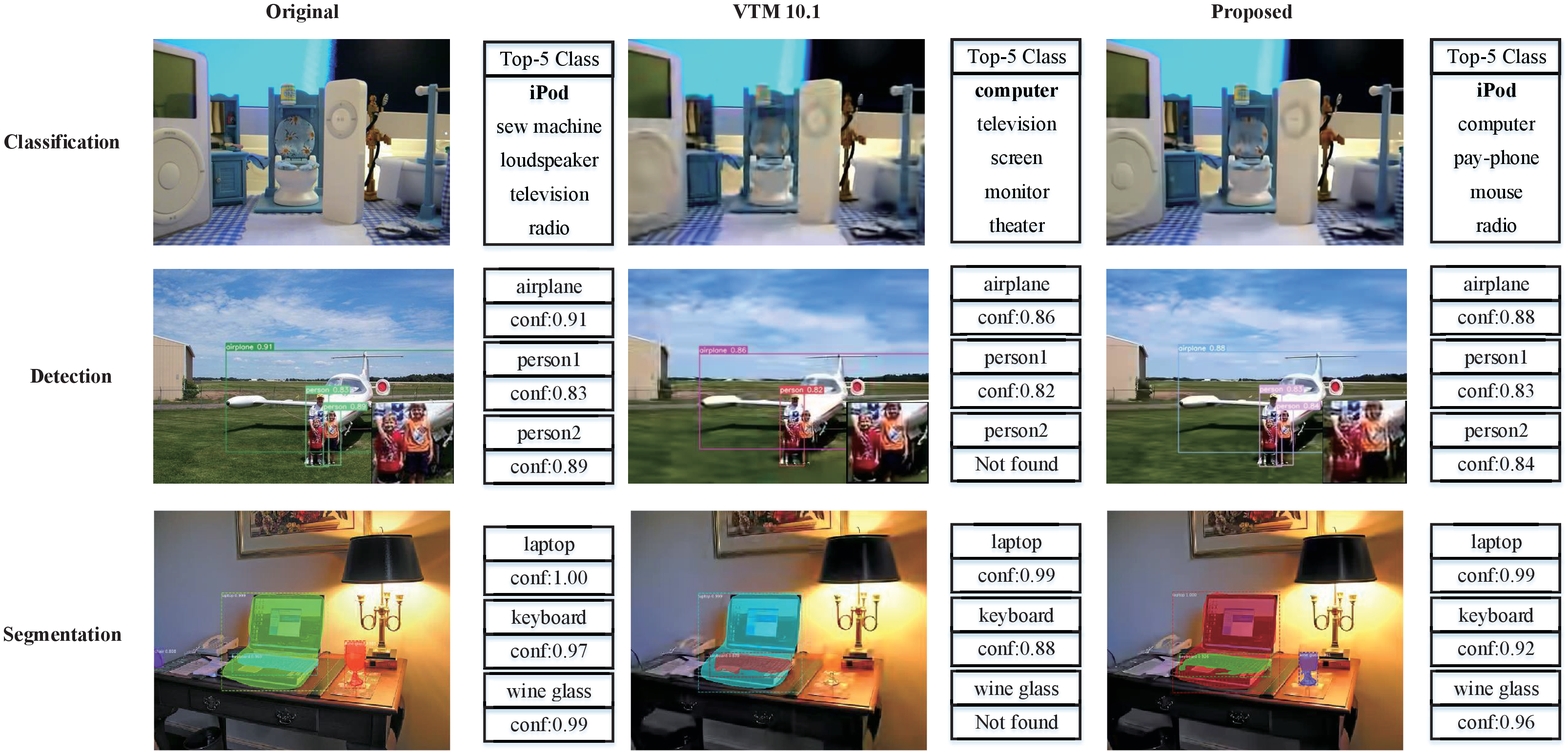}
	\caption{Comparison between the proposed model and VTM 10.1 on three visual analysis tasks. The first column is the original images. The second and the third column are the images encoded by VTM 10.1 and the proposed model in similar bitrate. The first row shows the performance on classifications. The label of the original image is iPod and the image encoded by VTM 10.1 is classified into computer category. As for the second and third rows, $conf$ denotes the confidences of the objects.}
	\label{fig:example}
\end{figure*}
\subsection{Performance Evaluation and Analysis}

As shown in Fig.~\ref{fig:curves}, the proposed model shows better performance on every visual task. For the classification task on ImageNet, the top5-accuracy of the proposed model with bpp equalling to 0.09 is lower than it of VTM 10.1. However, with bpp greater than 0.10, the accuracy of top-1 and top-5 of the proposed model is better than that of VTM 10.1. The difference between the performance of the proposed model and anchor is constant. Unlike classification, differences of mAP@50 between the proposed model and anchor on detection and segmentation tasks are variational: with the increase of the bitrate, the improvement of the proposed model is upgrading. This phenomenon reveals that the increase of bpp in low bitrate tends to improve the quality of high ROIM regions. In Fig.~\ref{fig:example}, some examples of different tasks are shown. For the classification task, the image coded by the proposed model could be classified correctly. As for the detection and segmentation tasks, the confidences of each object coded by the proposed model is increasing. Some objects, such as person2 in the detection example and wine glass in the segmentation example, can not be detected or segmented if they are encoded by VTM 10.1. However, in the third column, more bits are allocated to regions with high degrees of importance for machine vision by the proposed model. As a result, all of the objects encoded by the proposed model could be detected or segmented in Fig.~\ref{fig:example}. 
\par The Bjontegaard method is deployed for our performance evaluation. The most common distortion measurement of BD-Rate applied in image coding is PSNR. To compare the performance, classification accuracy and mAP@50 are taken as the distortion to calculate BD-rate. and the BD-Rate is listed in Table~\ref{tab:SB:2}. According to the experimental results, the proposed framework could improve the performance of visual tasks. On three totally different visual tasks, our approach could achieve about 19\% to 28\% bit-rate reduction.
\begin{table}[h]\small
	\begin{center}
		\caption{Rate-Accuracy Performance with top-1 accuracy (Top-1 Acc.), top-5 accuracy (Top-5 Acc.), mAP@50 as distortion, respectively.} \label{tab:SB:2}
		\begin{tabular}{|c|c|}
			\hline
			Visual Task&BD-Rate\\ \hline
			Classification(Top-1 Acc.)&-19.88\%\\ 
			Classification(Top-5 Acc.)&-19.13\%\\
			Detection(mAP@50)&-26.25\%\\ 
			Segmentation(mAP@50)&-28.17\%\\ 
			\hline
		\end{tabular}
	\end{center}
\end{table}
\section{Conclusion}
In this paper, we propose a visual analysis-motivated rate-distortion model for image coding. A new concept named ROIM is proposed to evaluate the degree of importance for each CTU. Furthermore, we use the ROIM to guide the bit allocation strategy. After comparison and analysis, we propose a distortion measurement model based on a neural network. A novel MSFD distortion is proposed to deal with the lacking of context information. The experimental results show that the images coded by the proposed scheme perform better than VTM in classification, detection, and segmentation tasks. In the future, we will extend intra image coding to iter mode coding optimization and test our model on more visual analysis tasks.

\bibliographystyle{icme2021}\footnotesize
\bibliography{icme2021}

\begin{thebibliography}{10}

\bibitem{he2016deep}
Kaiming He, Xiangyu Zhang, Shaoqing Ren, and Jian Sun,
\newblock ``Deep residual learning for image recognition,''
\newblock in {\em Proceedings of the IEEE conference on computer vision and
  pattern recognition}, 2016, pp. 770--778.

\bibitem{deng2009imagenet}
Jia Deng, Wei Dong, Richard Socher, Li-Jia Li, Kai Li, and Li~Fei-Fei,
\newblock ``Imagenet: A large-scale hierarchical image database,''
\newblock in {\em 2009 IEEE conference on computer vision and pattern
  recognition}. Ieee, 2009, pp. 248--255.

\bibitem{chen2019learning}
Zhibo Chen and Tianyu He,
\newblock ``Learning based facial image compression with semantic fidelity
  metric,''
\newblock {\em Neurocomputing}, vol. 338, pp. 16--25, 2019.

\bibitem{zhang2019recent}
Jiaqi Zhang, Chuanmin Jia, Meng Lei, Shanshe Wang, Siwei Ma, and Wen Gao,
\newblock ``Recent development of avs video coding standard: Avs3,''
\newblock in {\em 2019 Picture Coding Symposium (PCS)}. IEEE, 2019, pp. 1--5.

\bibitem{sullivan2012overview}
Gary~J Sullivan, Jens-Rainer Ohm, Woo-Jin Han, and Thomas Wiegand,
\newblock ``Overview of the high efficiency video coding (hevc) standard,''
\newblock {\em IEEE Transactions on circuits and systems for video technology},
  vol. 22, no. 12, pp. 1649--1668, 2012.

\bibitem{CFP}
Andrew Segall, Vittorio Baroncini, Jill Boyce, Jianle Chen, and Teruhiko
  Suzuki,
\newblock ``Joint call for proposals on video compression with capability
  beyond {HEVC},''
\newblock {\em JVET document, JVET-H1002}, 2017.

\bibitem{toderici2017full}
George Toderici, Damien Vincent, Nick Johnston, Sung Jin~Hwang, David Minnen,
  Joel Shor, and Michele Covell,
\newblock ``Full resolution image compression with recurrent neural networks,''
\newblock in {\em Proceedings of the IEEE Conference on Computer Vision and
  Pattern Recognition}, 2017, pp. 5306--5314.

\bibitem{balle2018variational}
Johannes Ball{\'e}, David Minnen, Saurabh Singh, Sung~Jin Hwang, and Nick
  Johnston,
\newblock ``Variational image compression with a scale hyperprior,''
\newblock {\em arXiv preprint arXiv:1802.01436}, 2018.

\bibitem{wang2009mean}
Zhou Wang and Alan~C Bovik,
\newblock ``Mean squared error: Love it or leave it? a new look at signal
  fidelity measures,''
\newblock {\em IEEE signal processing magazine}, vol. 26, no. 1, pp. 98--117,
  2009.

\bibitem{egiazarian2006new}
Karen Egiazarian, Jaakko Astola, Nikolay Ponomarenko, Vladimir Lukin, Federica
  Battisti, and Marco Carli,
\newblock ``New full-reference quality metrics based on hvs,''
\newblock in {\em Proceedings of the Second International Workshop on Video
  Processing and Quality Metrics}, 2006, vol.~4.

\bibitem{li2011visual}
Zhicheng Li, Shiyin Qin, and Laurent Itti,
\newblock ``Visual attention guided bit allocation in video compression,''
\newblock {\em Image and Vision Computing}, vol. 29, no. 1, pp. 1--14, 2011.

\bibitem{hadizadeh2013saliency}
Hadi Hadizadeh and Ivan~V Baji{\'c},
\newblock ``Saliency-aware video compression,''
\newblock {\em IEEE Transactions on Image Processing}, vol. 23, no. 1, pp.
  19--33, 2013.

\bibitem{khanna2015perceptual}
Meera~Thapar Khanna, Karan Rai, Santanu Chaudhury, and Brejesh Lall,
\newblock ``Perceptual depth preserving saliency based image compression,''
\newblock in {\em Proceedings of the 2nd International Conference on Perception
  and Machine Intelligence}, 2015, pp. 218--223.

\bibitem{oh2012video}
Hyungsuk Oh and Wonha Kim,
\newblock ``Video processing for human perceptual visual quality-oriented video
  coding,''
\newblock {\em IEEE Transactions on Image processing}, vol. 22, no. 4, pp.
  1526--1535, 2012.

\bibitem{zhu2018spatiotemporal}
Shiping Zhu and Ziyao Xu,
\newblock ``Spatiotemporal visual saliency guided perceptual high efficiency
  video coding with neural network,''
\newblock {\em Neurocomputing}, vol. 275, pp. 511--522, 2018.

\bibitem{shi2020reinforced}
Jun Shi and Zhibo Chen,
\newblock ``Reinforced bit allocation under task-driven semantic distortion
  metrics,''
\newblock in {\em 2020 IEEE International Symposium on Circuits and Systems
  (ISCAS)}. IEEE, 2020, pp. 1--5.

\bibitem{li2014lambda}
Bin Li, Houqiang Li, Li~Li, and Jinlei Zhang,
\newblock ``$\lambda$ domain rate control algorithm for high efficiency video
  coding,''
\newblock {\em IEEE transactions on Image Processing}, vol. 23, no. 9, pp.
  3841--3854, 2014.

\bibitem{VTM10.1}
Jianle Chen, Yan Ye, and Seung~Hwan Kim,
\newblock ``Algorithm description for {V}ersatile {V}ideo {C}oding and {T}est
  {M}odel 10 ({VTM} 10),''
\newblock {\em JVET document, JVET-S2002}, 2020.

\end{thebibliography}
\end{document}